# Prospection and dispersal in metapopulations: a perspective from opinion dynamics models

Daniela Molas and Daniel Campos

**Abstract** Dispersal is often used by living beings to gather information from conspecifics, integrating it with personal experience to guide decision-making. This mechanism has only recently been studied experimentally, facilitated by advancements in tracking animal groups over extended periods. Such studies enable the analysis of the adaptive dynamics underlying sequential decisions and collective choices. Here, we present a theoretical framework based on the Voter Model to investigate these processes. The model, originally designed to study opinion or behavioral consensus within groups through imitation, is adapted to include the prospection of others' decisions as a mechanism for updating personal criteria. We demonstrate that several properties of our model—such as average consensus times and polarization dynamics—can be analytically mapped onto those of the classical Voter Model under simplifying assumptions. Finally, we discuss the potential of this framework for studying more complex scenarios.

## 1 Introduction

The scientific study of decision-making in living beings poses a significant challenge across multiple disciplines. Traditionally associated with psychology and sociology, this field has recently garnered growing interdisciplinary interest, driven by advancements in experimental techniques and theoretical insights from ecology, mathematics, and physics. These developments have enabled more integrative approaches to understanding decision-making processes. Notable examples include the study of

Daniela Molas
Physics Dept, Universitat Autonoma de Barcelona, 08193 Cerdanyola del Vallès, e-mail: daniela.molas@autonoma.cat

Daniel Campos
Physics Dept, Universitat Autonoma de Barcelona, 08193 Cerdanyola del Vallès, e-mail: daniel.campos@uab.cat

collective movement in animal groups through the frameworks of complexity science [1] and fluid dynamics [2, 3, 4], as well as the analysis of cooperative behavior among rational agents in evolutionary game theory scenarios, such as the prisoner's dilemma [5, 6], or via decentralized communication systems, as observed in eusocial species [7, 8].

A particularly important aspect within this context concerns how groups of individuals prospect for information from peers to guide their decisions during extended decision-making processes. Information sharing and prospection serve as essential mechanisms for fostering behavioral and/or opinion consensus. This is of intrinsic interest for understanding collective phenomena in ecological and social systems, including positive outcomes (e.g., cooperation) and negative consequences (e.g., the tragedy of the commons).

The Voter Model [9] is one of the most prominent mathematical frameworks for studying opinion consensus in such contexts. In its simplest form, the Voter Model assumes that individuals (voters) make sequential choices from a discrete set of options (e.g., political parties) by imitating the decisions of their peers. Despite its simplicity, the Voter Model and its numerous extensions [10, 11, 12, 13, 14, 15, 16, 17, 18, 19, 20] have attracted sustained theoretical interest over the past decades. This model plays a central role in the statistical physics of social systems [10, 11], providing a robust framework for exploring the dynamics of consensus formation in both natural and artificial populations.

## 1.1 The Voter Model in Biology

While copying and imitation are common mechanisms in living systems, most studies on the Voter Model have been restricted to theoretical contexts, with real-life applications remaining relatively unexplored [21]. This is somewhat surprising, as natural selection itself can be viewed as a copying mechanism, where different options (genotypes) spread and compete, leading to either consensus (survival of the fittest) or polarization (coexistence through specialization or resource partitioning).

Several factors may explain the limited application of Voter-like models to experimental biology:

(i) Most versions of the Voter Model are restricted to populations with a fixed number of individuals, which is often unrealistic in biological systems.

(ii) Opinion consensus in biological populations typically arises through local and dynamic interactions. However, spatially explicit versions of the Voter Model remain scarce. The Sznajd model provides some notable exceptions; see [22, 23] for details.

(iii) As highlighted in [21], consensus among sepparate individuals is not always meaningful in biological systems. Instead, population-level or site-level descriptions are more relevant, yet extending Voter Models to these scales is nontrivial.

Despite these challenges, Voter-like models have been successfully applied to genetic drift and evolutionary dynamics, starting with Moran's pioneering work [24] and extending to more recent studies [25, 26, 27, 28, 29, 30]. Applications to fields like behavioral and movement ecology, however, remain rare (but see [31, 32, 33, 34]).

Imitation and majority rules, central to many Voter Models, are recognized as powerful drivers of both individual and social decision-making in animal groups [35, 36, 37]. A particularly relevant case involves seasonal species that annually select breeding sites. In such cases, individual experience is often complemented by social information gathered from the decisions of conspecifics, a process known as "prospection" [38]. The effect of prospection on fitness and reproductive success has only recently been explored, as it requires tracking animal movements over large temporal and spatial scales. Some datasets, particularly for seabirds [39], have become available recently through advances in GPS technology and other tracking methods, allowing a notable expansion of the field.

In this work, we explore how Voter-like models can be adapted to study the role of prospection in metapopulation dynamics [40], where different sites or patches represent the options available to "voters." Breeding decisions based solely on prior experience correspond to "stubborn voters," who repeatedly choose the same option, while prospection introduces an imitation dynamic within the model.

Our primary objective is to investigate how prospection influences opinion consensus, a typical outcome in classical Voter Models. Specifically, we examine: (i) how the time to achieve consensus varies with prospection rates or efficiency, and (ii) the emergence of alternative outcomes, such as polarization or stalemates. In the case of polarization, individuals coexist within distinct subpopulations, each maintaining differentiated opinions or preferences. Consequently, the system ceases to evolve, ultimately leading to an absorbing state. On the other hand, stalemates refer to stationary scenarios where these subpopulations coexist and interact by exchanging individuals but without reaching a unified consensus so, unlike polarization, system fluctuations hinder the attainment of an absorbing state.

To this end, we propose a version of the Voter Model where individuals accumulate personal information through prospection (Fig. 1). Similar approaches incorporating personal information have recently been explored, revealing mechanisms that can promote polarized states [12, 16]. In our adaptation, individuals base their decisions on personal/internal criteria, which, in a biological context, could represent attributes such as proximity to food or predation pressure. These criteria are updated through prospection visits to neighboring patches. Spatially explicit effects are expected to play a significant role, and we analyze these under two idealized scenarios: global prospection (where individuals can prospect for any patch) and local prospection (restricted to neighboring patches).

The structure of this work is as follows. In Section 2, we introduce the Voter-like model with personal criteria and describe its parameters. Section 3 examines the effects of personal information in the mean-field case, where all patches are equally accessible (global prospection). Section 4 addresses spatially explicit effects by considering local prospection. Finally, in Section 5, we summarize the main findings and propose ideas for future research.

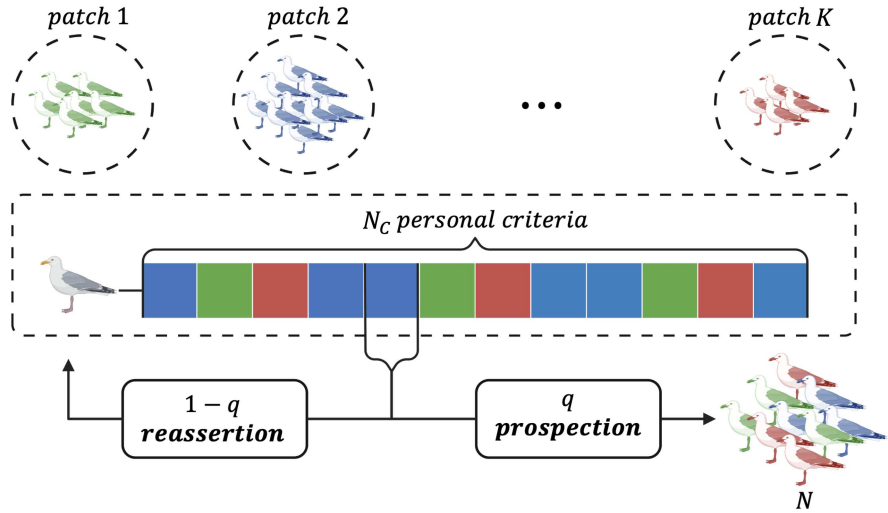

**Fig. 1** Model schematics. Each individual in the group has personal information based on $N_C$ criteria, each supporting a specific patch or breeding site (indicated by colors). Individual choices at each time step are influenced by these criteria, which are updated either by prospection (probability $q$) or by reinforcing the individual's current option (reassertion, probability $1 - q$). See the main text for further details.

## 2 Model

We consider a system of $N$ individuals which at each time step have to choose one from a set of $K$ different options available (with $K \geq 2$). In the classical Voter Model such decisions are simply taken by imitating what the other individuals do, though several variants exist which take into account different aspects of animal or human decision making. One such aspect, not much studied previously, is the idea that observing others' decisions might not necessarily lead to instantaneous imitation, but it just provides an external source of information thats adds to the background knowledge of the individual. To take into account this aspect, we assume that decisions are based on a set of $N_C$ personal criteria, which represent relevant features of the options. So that, the decision is expected to be the result of combining or prioritizing such features/criteria.

Let us illustrate this with the specific example of birds selecting a suitable patch annually for nesting (Fig. 1). To make their decision, the birds consider their individual preferences based on a set of $N_C$ criteria, such as proximity to food sources, predation risk, competition with conspecifics, availability of resting areas, etc.

Each individual has its own preferences (or personal information) regarding these criteria. If we denote individuals by the index $i = 1, 2, \ldots, N$, where $N$ is the total number of individuals, and criteria by the index $j = 1, 2, \ldots, N_C$, where $N_C$ is the total number of criteria considered, then the $j$-th criterion of the $i$-th individual, defined as $c_{i,j}$, can take possible values from $\{1, 2, \ldots, K\}$, where $K$ is the number of available options.

The values of these criteria are updated based on either prior experience or information acquired by the individual through exploring other patches. At each time step, (i) one criterion from each individual (say $i_1$) is chosen at random, (ii) a neighboring individual $i_2$ is selected randomly, and (iii) the chosen criterion $c_{i_1,j}$ is updated such that it adopts the option (patch) held by the neighboring individual $i_2$ with probability $q$, or with probability $1 - q$, it retains the option chosen by the individual $i_1$ itself. Hence, $q$ represents the probability that individual $i_1$ acquires new information from its neighbors, and updates its preferences accordingly; otherwise, the individual reaffirms its current preferences.

Next, following these personal preferences, the corresponding decisions are made as follows. At each time step, (i) we choose one individual at random, (ii) we choose one of its personal criteria at random, and (iii) the choice made by the individual is updated to the option supported by the selected criterion. So, the agent is assumed to be prioritizing such particular criterion/feature at that moment.

By introducing this dual dynamics (a dynamics for taking decisions, and another one for updating criteria) we dissociate what individuals "do" from what individuals "think". Of course, if criteria are updated at a rate fast enough then they will reflect the real options chosen by the population, and then individuals decisions and criteria will be aligned. But there are also mechanisms (e.g. noisy or uncertain prospection) that could lead the individuals to "behave" and "think" differently. While this is a very attractive possibility to explore, in the present contribution we will rather focus for simplicity on simple situations where individuals are expected on average to "think" and "behave" similarly.

We define $n_\alpha(t)$ as the amount of individuals which are holding option $\alpha$ at time step $t$, and $n_{\alpha,\beta}(t)$ as the total number of criteria for these individuals that give support to option $\beta$. As a result, $N = \sum_\alpha n_\alpha$ and $N_C = \sum_\beta n_{\alpha,\beta}$ hold, and the case $n_{\alpha,\alpha} = N_C$ would represent a situation in which all the criteria for the individuals holding option $\alpha$ give support to that option.

From the definitions above, we can define a set of *magnetizations* as

$$m_\alpha \equiv n_\alpha - \sum_{\beta \neq \alpha} n_{\alpha,\beta} = 2n_\alpha - N, \tag{1}$$

for $\alpha = 1, 2, \ldots, K$. This magnetization explicitly depends on the fraction of individuals holding option $\alpha$ and reflects the level of prevalence of this option over the others. Defined in this way, the consensus of the population, where all individuals support the same option $\alpha$, satisfies $m_\alpha = N$, and $m_{\alpha'} = -N$ for $\alpha' \neq \alpha$.

Following the usual guidelines from the classical Voter Model with multiple options (see, e.g., [14, 16]), a Master equation for the evolution of these magnetizations can be written in the mean-field approximation. By introducing $P(m_\alpha, t)$ as the probability that the magnetization $m_\alpha$ takes a particular value at time step $t$, the Master equation reads

$$P(m_\alpha, t+1) = \sum_{\beta \neq \alpha}^{K} \left[ \frac{n_\beta}{N} \frac{n_{\alpha,\beta}}{N_C} P(m_\alpha - 2, t) + \frac{n_\alpha}{N} \frac{n_{\beta,\alpha}}{N_C} P(m_\alpha + 2, t) \right] +$$

$$+ \left( \frac{n_\alpha}{N} \frac{n_{\alpha,\alpha}}{N_C} + \sum_{\beta \neq \alpha} \frac{n_\beta}{N} \frac{\sum_{\gamma \neq \alpha} n_{\gamma,\beta}}{N_C} \right) P(m_\alpha, t). \tag{2}$$

Let us analyze each term one by one:

1. The first term represents the probability that an individual holding option $\beta$ is chosen at random at time $t$ ($n_\beta/N$), and that the criterion assigned randomly supports option $\alpha$ ($n_{\alpha,\beta}/N_C$), causing the individual to switch from $\beta$ to $\alpha$. As a result, $n_\alpha$ increases by 1, which causes $m_\alpha$ to increase by 2.
2. The second term represents the possibility that the individual chosen at random holds option $\alpha$, but the assigned criterion supports option $\beta$, causing the individual's option to change from $\alpha \longrightarrow \beta$. As a result, $n_\alpha$ decreases by 1, which causes $m_\alpha$ to decrease by 2.
3. The third term stands for the situation where individuals do not switch their option. The expression $\frac{n_\alpha}{N} \frac{n_{\alpha,\alpha}}{N_C}$ represents the case where the option held by the individual and that supported by the criteria are both $\alpha$, so nothing changes. Similarly, $\sum_{\beta \neq \alpha} \frac{n_\beta}{N} \frac{\sum_{\gamma \neq \alpha} n_{\gamma,\beta}}{N_C}$ stands for those cases where both the option of the individual and the option supported by the criterion are different from $\alpha$, for which the value of $m_\alpha$ remains the same.

In our particular model, the imitation rule is not explicitly included in the dynamics above of how the individuals take options, but on how the criteria are updated. For this reason, it is more convenient and instructive to work with *magnetizations* defined at the level of the criteria. These can be defined as

$$\sigma_\alpha \equiv n_{\alpha,\alpha} - \sum_{\beta \neq \alpha} n_{\alpha,\beta} \tag{3}$$

since the criteria are part of a set of options. It follows that $\sigma_\alpha = N_C$ whenever $n_{\alpha,\alpha} = N_C$, which implies that the set of criteria is aligned with the same option $\alpha$.

The Master equation associated with $\rho(\sigma_\alpha, t)$, which represents the probability density of $\sigma_\alpha$ at time $t$, takes a form similar to that above for $P(m_\alpha, t)$. However, this equation consists of two distinct terms: the first represents the reassertion mechanism, with probability $(1-q)$, and the second describes the dynamics induced by imitation rules, with complementary probability $q$:

$$\rho(\sigma_\alpha, t+1) = (1-q) \sum_{\beta \neq \alpha}^{K} \left[ \frac{n_\beta}{N} \frac{n_{\alpha,\beta}}{N_C} \rho(\sigma_\alpha - 2, t) + \frac{n_\alpha}{N} \frac{n_{\beta,\alpha}}{N_C} \rho(\sigma_\alpha + 2, t) \right] +$$

$$+ q \sum_{\beta \neq \alpha} \left[ \frac{n_\beta}{N} \frac{\sum_{\gamma=1}^{K} n_{\alpha,\gamma}}{K N_C} \rho(\sigma_\alpha - 2, t) + \frac{n_\alpha}{N} \frac{\sum_{\gamma=1}^{K} n_{\beta,\gamma}}{K N_C} \rho(\sigma_\alpha + 2, t) \right] +$$

$$+ \left( \frac{n_\alpha}{N} \frac{n_{\alpha,\alpha}}{N_C} + \sum_{\beta \neq \alpha} \frac{n_\beta}{N} \frac{\sum_{\gamma \neq \alpha} n_{\gamma,\beta}}{K N_C} \right) \rho(\sigma_\alpha, t), \tag{4}$$

where the quotient $\frac{\sum_{\gamma=1}^{K} n_{\alpha,\gamma}}{K N_C}$ represents the fraction of criteria supporting option $\alpha$ for the whole population, and similarly, $\frac{\sum_{\gamma=1}^{K} n_{\beta,\gamma}}{K N_C}$ represents the fraction of criteria supporting option $\beta$. Note that in the particular case $K = 2$ (only two options available) the previous expression simplifies to

$$\rho(\sigma, t+1) = (1-q) \left[ \frac{n_2}{N} \frac{n_{1,2}}{N_C} \rho(\sigma - 2, t) + \frac{n_1}{N} \frac{n_{2,1}}{N_C} \rho(\sigma + 2, t) \right] +$$

$$+ q \left[ \frac{n_2}{N} \frac{n_{1,1} + n_{1,2}}{2 N_C} \rho(\sigma - 2, t) + \frac{n_1}{N} \frac{n_{2,1} + n_{2,2}}{2 N_C} \rho(\sigma + 2, t) \right] +$$

$$+ \left( \frac{n_1}{N} \frac{n_{1,1}}{N_C} + \sum_{\beta \neq \alpha} \frac{n_\beta}{N} \frac{\sum_{\gamma \neq \alpha} n_{\gamma,\beta}}{2 N_C} \right) \rho(\sigma, t), \tag{5}$$

with $\sigma \equiv (n_{2,1} + n_{2,2}) - (n_{1,1} + n_{1,2})$. While this equation does not admit a simple solution, this simplified case allows us to gain some understanding about the effect that each mechanism plays on the dynamics of the system. For this, in the following we consider separately the effect of the prospection and the reassertion mechanisms.

## 2.1 Adiabatic approximation

Before interpreting the system dynamics, let us introduce a key concept to understand the arguments and conclusions that we present below. As mentioned, our model distinguishes between what individuals "do" (the options they take) and what they "think" (the options supported by their criteria). When imitation is highly effective and criteria are updated rapidly, these criteria faithfully reflect the options chosen by individuals in the group. In this situation, individuals' decisions mimic what others are doing, as in the classical Voter Model. We refer to this situation as an *adiabatic approximation*, as it allows us to separate the timescale of criteria updates (fast scale) from that of individual decisions (slow scale). From this perspective, the classical Voter Model can be seen as an adiabatic approximation of our model with personal criteria. Mathematically, this approximation is expressed as:

$$\frac{n_\alpha}{N} = \frac{n_{\alpha,\beta}}{N_C} \tag{6}$$

for any $\alpha$ and $\beta$.

In this context, the statistics of our model can be directly derived from the classical model whenever the adiabatic approximation holds. In particular, the times required to reach consensus or for an option to disappear in the classical model with multiple options are defined as follows:

- The consensus time, represented by $T_c$, is defined as the time it takes for all individuals within the system to reach the same option ($n_\alpha = N$ for some $\alpha$).

- Extinction time, represented by $T_e$, is defined as the time required for one of the options to disappear completely from the system ($n_\alpha = 0$ and $n_{i,\alpha} = 0$ for any $i$, for option $\alpha$).

Under these definitions, the average (computed over the different stochastic realizations of the system) of the consensus and extinction times in the classical model is known to satisfy [15]:

$$\langle T_c \rangle = \frac{K-1}{K}N \qquad \langle T_e \rangle = \frac{N}{K(K-1)}, \tag{7}$$

where $K$ is the number of available options and $N$ the total number of individuals. Under simplified conditions, we will verify in the following Sections how these exact results also apply to our model with personal criteria.

## 2.2 Prospection mechanism (case $q = 1$)

We consider the case $q = 1$ in Eq. (5). Using a Kramers-Moyal expansion in time and space, one obtains the following Fokker-Planck equation:

$$\frac{\partial \rho}{\partial t} = \left[ \frac{n_{2,1} + n_{2,2} - n_{1,1} - n_{1,2}}{NN_C} \right] \frac{\partial \rho}{\partial \sigma} + \left[ \frac{(n_{2,1} + n_{2,2})^2 + (n_{1,1} + n_{1,2})^2}{2NN_C} \right] \frac{\partial^2 \rho}{\partial \sigma^2}, \tag{8}$$

where the explicit dependences in $\rho(\sigma, t)$ have been omitted for ease of notation.

After some algebraic manipulations, it is possible to rewrite this equation in terms of $\sigma$ and $m \equiv n_2 - n_1$, leading to:

$$\frac{\partial \rho}{\partial t} = \left[ \frac{m - \sigma}{NN_C} \right] \frac{\partial \rho}{\partial \sigma} + \left[ 1 - \frac{m\sigma}{NN_C} \right] \frac{\partial^2 \rho}{\partial \sigma^2}. \tag{9}$$

From this, the effects of the drift and diffusion terms in (9) on the system can be analyzed. The drift term drives the system toward the condition $m = \sigma$, aligning the options with the criteria. This effectively brings the system into a regime where the adiabatic approximation (6) holds.

The diffusion term determines the magnitude of the fluctuations around the condition $m = \sigma$, as promoted by the drift term. Similarly to the classical Voter

Model, cases with vanishing magnetization ($m \approx \sigma \approx 0$) produce larger fluctuations. Conversely, as consensus is approached, with either $m \approx N, \sigma \approx N_C$ or $m \approx -N, \sigma \approx -N_C$, fluctuations tend to diminish. Consequently, system fluctuations favor departures from the homogeneous state $m = 0$, where both options 1 and 2 coexist, and promote the consensus state where only one option prevails, resulting in extreme magnetization values.

### 2.3 Reassertion mechanism (case $q = 0$)

By carrying out the same analysis as in the previous section (using the Kramers-Moyal expansion), the resulting Fokker-Planck equation for the dynamics of $m$ is given by:

$$\frac{\partial \rho}{\partial t} = \left[ \frac{n_{2,1} - n_{1,2}}{N N_C} \right] \frac{\partial \rho}{\partial \sigma} + \left[ \frac{(n_{2,1} + n_{2,2})^2 + (n_{1,1} + n_{1,2})^2}{2 N N_C} \right] \frac{\partial^2 \rho}{\partial \sigma^2}. \tag{10}$$

In this case, it is not possible to express the drift and diffusion coefficients in terms of $m$ and $\sigma$, as the update of an individual's option depends solely on their personal criteria, which introduces local effects into the model.

From (10), we observe that the drift term has a rather neutral effect, promoting the symmetry condition $n_{2,1} = n_{1,2}$ between the two options, which aligns with the adiabatic approximation. Meanwhile, the diffusion term remains the same as in Eq. (9), ensuring that consensus is eventually reached through random changes in individuals' decisions driven by imitation (alternatively, consensus can also be influenced by an imbalance in the initial conditions or in the update rules, favoring one of the two options).

In conclusion, the dynamics of the system are largely governed by the imitation dynamics in the criteria update, which promote the emergence of the adiabatic approximation conditions if such imitation effects are sufficiently fast or strong. In the next section, we explore these ideas numerically.

## 3 Global prospection (mean-field case)

In previous studies [16], two possible outcomes were identified in the voter model with personal information: consensus and population polarization. Our main objective is to identify possible mechanisms by which these two outcomes can emerge or be promoted, and to analyze their characteristic timescales. Exact analytical solutions are generally unattainable for our model, so we focus on numerical results. However, in cases where the adiabatic approximation mentioned above is fulfilled, we find that most of our results can be explained using the classical Voter Model.

If the imitation rule allows all criteria to be updated by imitating any individual in the population, the only possible equilibrium state of the system is *consensus*, defined by

$$n_\beta = \delta_{\beta\alpha} N \qquad n_{\beta,\gamma} = \delta_{\beta\alpha} \delta_{\gamma\alpha} N_C, \tag{11}$$

where $\delta_{\alpha,\beta}$ represents the Kronecker delta function, and $\alpha$ is the option reached after consensus. Essentially, condition (11) implies that (i) all individuals in the system share the same option, and (ii) all their criteria support that option.

One of the most interesting questions concerns how the different parameters in the system affect the consensus time, $T_c$, required to reach this unique equilibrium state. For multi-option imitation models, consensus is reached through a cascade of events [14], in which one of the $K$ available options first disappears (as soon as there are no criteria supporting it), leaving the system in a steady situation. Subsequently, a second option disappears, and so on, until only one option remains. The *first extinction time*, $T_e$, refers to the time required for the first of the $K$ initial options to disappear completely from the system, thereby reducing it to a $(K-1)$-option model. In other words, it marks the first event in which an option becomes entirely absent from the system. This quantity is of particular interest as it represents the onset of the cascade of option eliminations that eventually leads to consensus.

In Figure 2, we present numerical results for the mean-field case, showing both the average first extinction times (left column) and consensus times (right column) for different values of $K$ (the number of available options) and imitation probabilities $q$, as a function of the number of personal criteria $N_C$. In this case, instead of considering local interactions between neighbors, it is assumed that each individual is equally influenced by the global average of opinions in the population.

The results reveal that in the regime $N_C \ll N$, extinction and consensus dynamics are largely governed by prospection. This is evident as the average times for both processes are inversely proportional to $q$ in this regime. When imitation operates at a fast rate, it accelerates the consensus dynamics and promotes convergence to the equilibrium state at a rate proportional to the imitation rate.

For $N_C \gg N$, a more complex effect emerges. In this case, extinction and consensus times converge slowly for all values of $q$ (complete convergence is not observed numerically in Figure 2, as computation times scale with $N_C^2$, limiting simulations for large $N_C$). Here, imitation dynamics alone is insufficient to achieve consensus; it also requires alignment between individual options and the criteria. While this alignment is nearly instantaneous for small $N_C$, it requires significantly long times for large $N_C$, introducing a new timescale into the model. Specifically, when $q$ is large, imitation is faster but this causes criteria to change frequently, slowing the alignment process. Conversely, small $q$ values slow imitation but accelerate alignment.

We find that first extinction dynamics occur more rapidly in systems with a larger number of options, as expected from (7). This confirms that extinction of each particular option occurs with independent probability, meaning that the greater the number of options, the shorter the time required for at least one to become extinct.

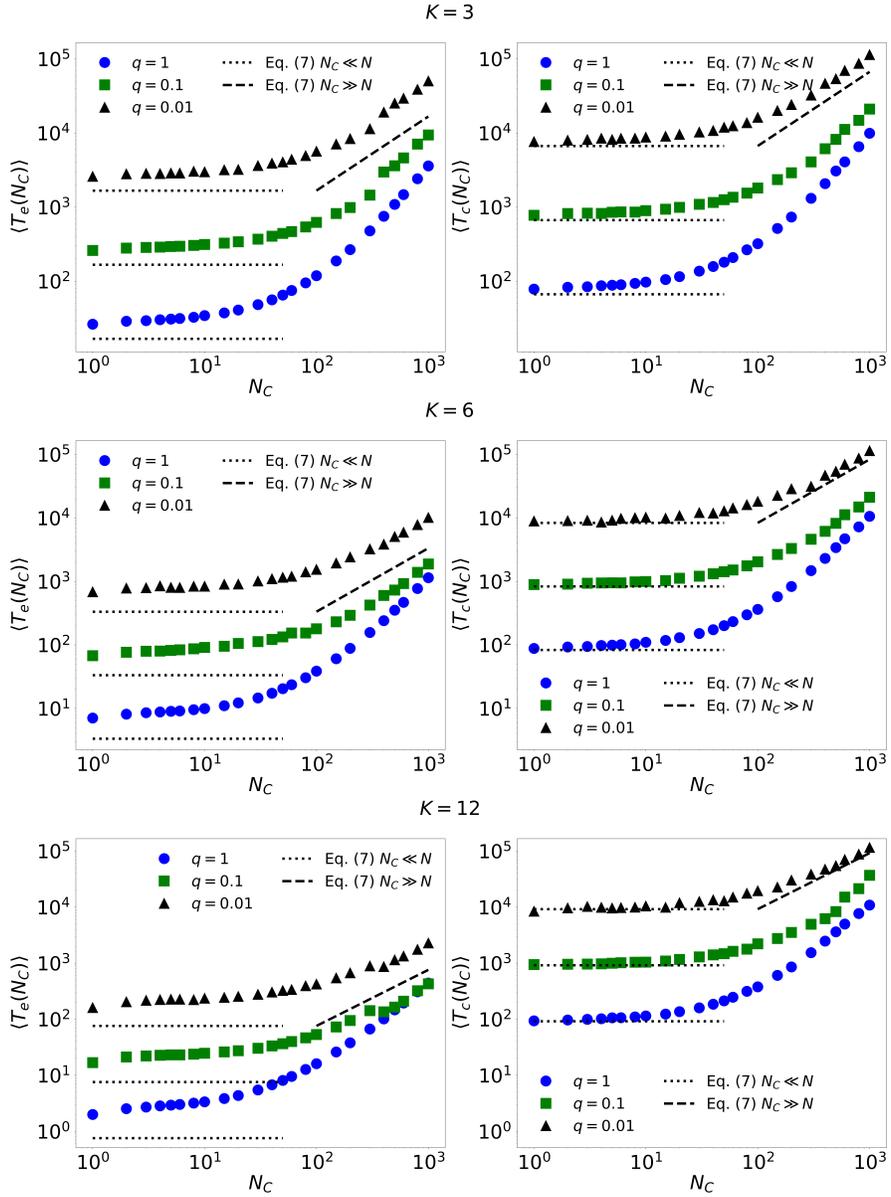

**Fig. 2** Average extinction (left) and consensus (right) times as a function of the number of criteria $N_C$. Dashed and horizontal dotted lines correspond to predictions from the classical model in Eq. (7) after parameter reinterpretation (see main text for details).

Finally, we observe two distinct regimes influenced by the number of criteria, $N_C$, corresponding to the conditions $N_C \ll N$ and $N_C \gg N$. In the former, consensus for all criteria of a single individual is achieved much faster than consensus among individuals. The bottleneck for reaching the equilibrium condition (11) is therefore achieving consensus among individuals. This implies that while criteria for a single individual can align quickly, the overall alignment of all individuals becomes the limiting factor. In other words, the time required for the entire population to agree on a single option dictates the pace of the system's approach to equilibrium. As a result, the classical prediction (7) applies in this regime (see horizontal dotted lines). Furthermore, extinction and consensus times remain nearly constant in this regime, though numerical simulation data admit power-law fits for $\langle T_e \rangle \sim N_C^{h_1}$ and $\langle T_c \rangle \sim N_C^{h_2}$, with exponents $h_1$ and $h_2$ taking values between 0 and 0.1 for the range of parameters explored here.

For the regime $N_C \gg N$, the bottleneck for reaching consensus is determined by the second condition in (11). In this case, the classical result (7) still applies, but $N$ should be replaced by $N \times N_C$ (the total number of criteria in the system), so both extinction and consensus times satisfy the scaling $\langle T \rangle \sim N \cdot N_C$.

Interestingly, modifying the rate at which criteria are updated does not significantly affect first extinction and consensus times, provided the imitation probability $q$ remains constant. To test this, we modified the model algorithm so that at each time step, only a fraction $f$ of the personal criteria of each individual were updated on average. The initial model corresponds to $f = 1/N_C$ (i.e., only one criterion is updated per time step). The value of $f$ has a minor impact on extinction and consensus dynamics for both $N_C \ll N$ and $N_C \gg N$ (not shown). This confirms that equilibrium is governed primarily by the imitation rate $q$. Rapid updates of personal criteria are ineffective unless imitation is sufficiently effective.

## 4 Local prospection

In this section, we address the question: "How can prospection dynamics be modified to allow the coexistence of multiple options?" Note that complete consensus is the only possible state under global prospection. However, from an ecological perspective, consensus can lead to practical issues, such as the tragedy of the commons or inefficient resource exploitation, due to the entire population converging on a single option. In contrast, the stable coexistence of options or species, referred to as polarization in the context of the Voter Model, requires alternative mechanisms to counteract imitation and prospection dynamics, as explored in [16].

As a potential mechanism to promote polarization, we consider that in real systems, prospection predominantly occurs within a local context. For example, birds or other animals primarily prospect neighboring patches to extract information, while visits to distant patches are infrequent or nonexistent. To simplify, we consider a scenario where options (or breeding places) are arranged in a one-dimensional lat-tice (labeled again from 1 to $K$), and prospection is restricted to consecutive or neighboring places.

Restricting prospection to local vicinity causes the imitation dynamics to lead to random extinction of different options at different regions. Consequently, the system can fragment into small subgroups of patches isolated from each other, with consensus dynamics operating independently within each subgroup. Over time, only a discrete set of isolated options survive; this corresponds to the *polarized* state.

One key question is how the number of available options, $K$, influences polarization dynamics. When fewer options are available, isolated populations are less likely to form because the available options are strongly connected, allowing a single option to dominate and absorb the others, resulting in consensus. Conversely, as $K$ increases, the probability of isolated populations emerging becomes significantly higher.

Figure 3 illustrates the numerical results, showing how the probability of reaching polarization (i.e., stable coexistence of isolated subpopulations) exhibits a phase transition for $K$ between approximately 6 and 8. Remarkably, this behavior is largely independent of the model parameters ($q$, $N_C$). It is primarily driven by the number of available options and the likelihood of isolation during the cascade of successive extinctions. This confirms that local prospection alone can drive the system toward polarized states. This finding opens avenues for exploring more realistic dynamics in future work, such as prospection governed by specific dispersal kernels, to better understand metapopulation dynamics and coexistence in real habitats or environments.

For completeness, Figure 4 shows the distribution of subpopulation sizes after the system reaches its final state. In cases of consensus, the size always corresponds to $N$, whereas in polarized states, different sizes are observed. The size distribution exhibits a peak at $N$ that decreases as $K$ increases, indicating that polarization becomes more likely. For sufficiently large $K$, the population fragments into smaller subpopulations, and the distribution skews toward smaller sizes.

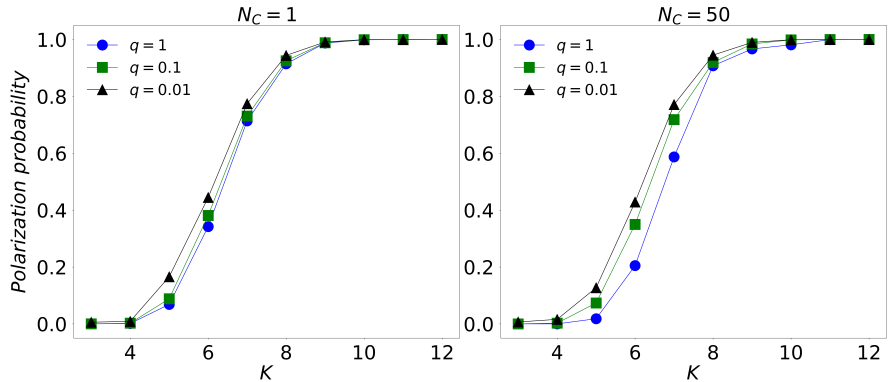

**Fig. 3** Polarization probability as a function of the number of options $K$. All data were obtained by considering $N = 100$ individuals and averaging over 10000 realizations of the process.

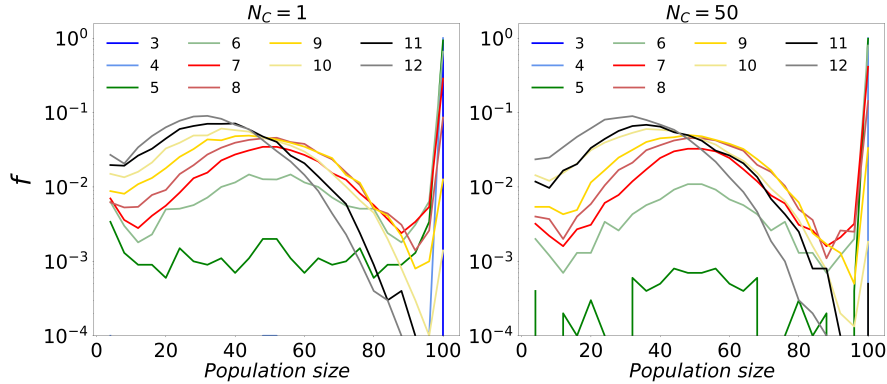

**Fig. 4** Population size distribution. Data were obtained by considering $N = 100$, $q = 1$, and averaging over 10000 realizations of the process.

Finally, in Figure 5, we examine how extinction times decay with the number of options $K$, confirming that the extinction of any option occurs with constant and independent probability. This dependence is related to the number of connections between patches. In the one-dimensional lattice considered, the number of initial connections is $K - 1$. Thus, the extinction time scales approximately as $\langle T_e \rangle \sim \langle T_e \rangle^* / (K-1)$, where $\langle T_e \rangle^*$ represents the extinction time for $K = 2$. This theoretical prediction, represented by dashed lines in Figure 5, aligns well with numerical results, particularly in the thermodynamic limit where $N$, $N_C$, and $K$ are very large, ensuring that the adiabatic approximation holds.

The colors in the figure correspond to different values of $q$, while the parameter $f$ represents the fraction of personal criteria updated at each time step, expressed as a value between 0 and 1.

## 5 CONCLUSIONS

In this work, we have introduced an extension of the classical Voter Model by incorporating the concept that prospecting information from others does not result in immediate imitation but instead contributes to an accumulation process that eventually leads to such behavior. Specifically, we considered a dual dynamics framework in which individuals make decisions based on internal personal criteria, while these criteria are updated through an imitation rule akin to that in the classical model. The dynamics of these criteria act as an intermediate process that delays the system's convergence to consensus, requiring alignment between individuals' actual choices and their criteria.

We have demonstrated that our model with personal criteria can be effectively understood in terms of the classical Voter Model, provided the alignment process occurs rapidly and does not become a significant bottleneck for consensus. This scenario corresponds to the *adiabatic approximation*, which we have used as a simplified regime to illustrate the model's applicability. Extending the model to the multi-option case ($K > 2$), we have also shown how a local prospection mechanism can naturally lead to polarized states, characterized by the coexistence of isolated populations reaching each a particular consensus locally. The existence of this polarized regime is seen to be essentially a consequence of the number of options available, $K$, while the other parameters in the model ($q, N_C$) simply rescale the time required to reach that state.

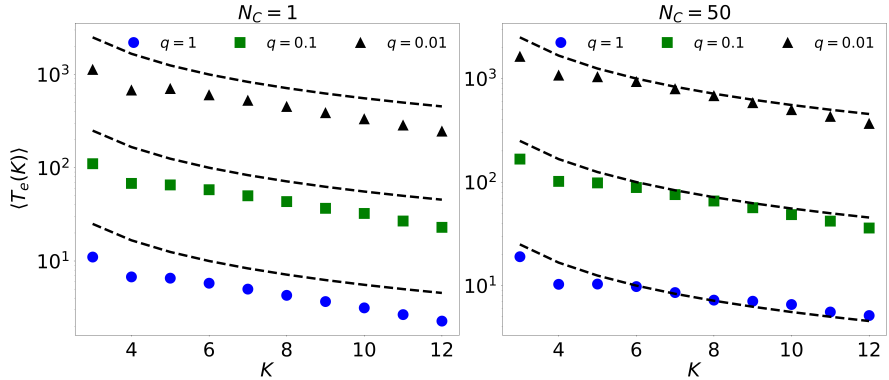

**Fig. 5** Average extinction time as a function of the number of options $K$. Dashed lines represent the theoretical approximation $\langle T_e \rangle \sim \langle T_e \rangle^*/(K-1)$ (see main text for details). The colors correspond to different values of $K$, while the parameter $f$ represents the fraction of personal criteria updated at each time step, expressed as a value between 0 and 1. All data were obtained by considering $N = 100$ individuals and averaging over 10000 realizations of the process.

Future work should explore scenarios where the duality between (i) decision-making and (ii) criteria updating becomes more complex and less straightforward, potentially breaking the adiabatic approximation, as a consequence of memory effects, biased/persistent dynamics in the choice of the criteria, etc. This could lead to intriguing phenomena, such as populations that "act" and "think" differently, or scenarios where consensus is unattainable due to noisy or uncertain prospection. Such cases represent promising avenues for further research.

Additionally, experimental testing and calibration of the model appear feasible using large-scale tracking datasets (e.g., GPS) for seasonal birds, as in [38, 39]. In this context, the criteria in the model could be linked to relevant fitness metrics, and individuals' visits to neighboring patches could serve as a measurable proxy for prospection. These extensions illustrate the broader applicability and potential interest of the approach presented here.

**Acknowledgements** DC acknowledges the organizers of the Isaac Newton Institute program on *Mathematics of movement: an interdisciplinary approach to mutual challenges in animal ecology and cell biology* for their kind invitation to the program and for discussions, which have substantially contributed to the present work. DM acknowledge financial support from the Coordination of the National Scholarship Program Carlos Antonio López (BECAL). The Authors acknowledge the financial support of the Spanish government under grant PID2021-122893NB-C22.

**Competing Interests** The authors have no conflicts of interest to declare that are relevant to the content of this chapter.

# References

1. Ioannou C.C., Laskowski K.L. (2023) *A multi-scale review of the dynamics of collective behaviour: from rapid responses to ontogeny and evolution*. Phil. Trans. Roy. Soc. B 378:20220059.
2. Vicsek T., Czirók A, Farkas I.J, Helbing D. (1999) *Application of statistical mechanics to collective motion in biology*. Physica A 274, 182-189.
3. Fodor E, Marchetti M.C. (2018) *The statistical physics of active matter: From self-catalytic colloids to living cells*. Physica A 504, 106-120.
4. Ouellette N.T. (2019) *The Most Active Matter of All*. Matter 2, 297.299.
5. Roca C.P., Cuesta J.A., Sánchez A. (2009) *Evolutionary game theory: Temporal and spatial effects beyond replicator dynamics*. Phys. Life Rev. 6(4), 208-249.
6. McAvoy A., Allen B., Nowak M.A. (2020) *Social goods dilemmas in heterogeneous societies*. Nature Hum. Behav. 4(8), 819-831.
7. Gordon D.M. (2010) *Ant encounters: Interaction Networks and Colony Behavior*. Cambridge Univ. Press, Cambridge.
8. Piñero J., Solé R. (2019) *Statistical physics of liquid brains*. Phil. Trans. R. Soc. B 374:20180376.
9. Holley, R. A., Liggett, T. M.(1975). *Ergodic theorems for weakly interacting infinite systems and the voter model*. The annals of probability, 643-663.
10. Redner, S. (2019). *Reality-inspired voter models: A mini-review*. Comptes Rendus Physique, 20(4), 275-292.
11. Jedrzejewski A., Sznajd-Weron, K. (2019) *Statistical Physics Of Opinion Formation: Is it a SPOOF?* Comptes Rendus Physique, 20(4), 244-261.
12. G. De Marzo, A. Zaccaria, C. Castellano. (2020) *Emergence of polarization in a voter model with personalized information*. Phys. Rev. Res. 2, 043117.
13. Liggett, T. M. (1985). *Interacting particle systems (Vol. 2)*. New York: Springer.
14. Ramirez, L., San Miguel, M., Galla, T. , (2022) *Local and global ordering dynamics in multistate voter models*. (Phys. Rev. E, 106(5): 054307).
15. Starnini, M., Baronchelli, A., Pastor-Satorras, R. (2012). *Ordering dynamics of the multi-state voter model*. J. Stat. Mech.: Theor. Exp., 2012(10), P10027.
16. Iannelli, G., De Marzo, G., Castellano, C. (2022). *Filter bubble effect in the multistate voter model*. Chaos, 32(4), 043103.
17. Littler, R. A. (1975). *Loss of variability at one locus in a finite population*. Math. Biosci., 25(1-2), 151-163.
18. Kimura, M. (1955). *Random genetic drift in multi-allelic locus*. Evolution, 419-435.
19. Mobilia, M. (2003). Does a single zealot affect an infinite group of voters?. Phys. Rev. Lett., 91(2), 028701.
20. Herrerías-Azcué, F., Galla, T. (2019). Consensus and diversity in multistate noisy voter models. Phys. Rev. E, 100(2), 022304.
21. B. Oborny. (2023) *Lost in translation? - Caveat to the application of the voter model in ecology and evolutionary biology*. Sci. Prog. 106(2):368504231175324.
22. K. Sznajd-Weron, J. Sznajd. (2000) *Opinion evolution in closed community*. Int. J. Mod. Phys. C 11(6), 1157–1165.

23. K. Sznajd-Weron, J. Sznadj, T. Weron. (2021) *A review on the Sznajd model — 20 years after*. Physica A 565, 125537.

24. P.A.P. Moran. (1958) *Random processes in genetics*. Math. Proc. Camb. Philos. Soc. 54: 60–71.

25. G.W.A. Constable, A.J. McKane. (2015) *Stationary solutions for metapopulation Moran models with mutation and selection*. Phys. Rev. E 91(3), 032711.

26. G.W.A. Constable, A.J. McKane. (2015) *Models of Genetic Drift as Limiting Forms of the Lotka-Volterra Competition Model*. Phys. Rev. Lett. 114(3), 038101.

27. D.M. Schneider, A.B. Martins, M.A.M. de Aguiar. (2016) *The mutation–drift balance in spatially structured populations*. J. Theor. Biol. 402, 9-17.

28. C. Molina, D.J.D. Earn. (2018) *The mutation-drift balance in spatially structured populations*. J. Math. Biol. 76(3), 645-678.

29. L. Heinrich, J. Müller, A. Teller, D. Zivkovic. (2018) *Effects of population- and seed bank size fluctuations on neutral evolution and efficacy of natural selection*. Theor. Pop. Biol. 123, 45-69.

30. J. Svoboda, S. Joshi, K. Chatterjee. (2024) *Amplifiers of selection for the Moran process with both Birth-death and death-Birth updating*. PLoS Comput Biol 20(3): e1012008.

31. J. Chave. (2001) *Spatial patterns and persistence of woody plant species in ecological communities*. Am. Nat. 157(1), 51-65.

32. J. Hidalgo, S. Suweis, A. Maritan. (2017) *Species coexistence in a neutral dynamics with environmental noise*. J. Theor. Biol. 413, 1-10.

33. C. Tu, S. Suweis, J. Grilli, M. Formentin, A. Maritan. (2019) *Reconciling cooperation, biodiversity and stability in complex ecological communities*. Sci. Rep. 9, 5580.

34. R. Martínez-García, C. López, F. Vázquez. (2021) *Species exclusion and coexistence in a noisy voter model with a competition-colonization tradeoff*. Phys. Rev. E 103, 032406.

35. C. Wang et al. (2020) *Decision-making process during collective movement initiation in golden snub-nosed monkeys (Rhinopithecus roxellana)*. Sci. Rep. 10, 480.

36. C. Winklmayr, A.B. Kao, J.B. Bak-Coleman, P. Romanczuk. (2020) *The wisdom of stalemates: consensus and clustering as filtering mechanisms for improving collective accuracy*. Proc. R. Soc. B 287:20201802.

37. H. Rajendran, A. Haluts, N.S. Gov, O. Feinerman. (2022) *Ants resort to majority concession to reach democratic consensus in the presence of a persistent minority*. Curr. Biol. 32(3), 645-653.

38. D. Oro, J. Becares, F. Bartumeus, J.M. Arcos. (2021) *High frequency of prospecting for informed dispersal and colonisation in a social species at large spatial scale*. Oecologia 197, 395-409.

39. J. Kralj et. al. (2023) *Active breeding seabirds prospect alternative breeding colonies*. Oecologia 201, 341-354.

40. I. Hanski. (1998) *Metapopulation dynamics*. Nature 396, 41-49.